\documentclass[useAMS,usenatbib]{mn2e}
\usepackage[dvips]{graphicx}

\title[Clustering of Radio Galaxies and Quasars]{Clustering of Radio Galaxies
and Quasars}
\author[Donoso et al.]{E.
Donoso$^{1}$\thanks{E-mail:edonoso@mpa-garching.mpg.de}, Cheng Li$^{1,2}$, G.
Kauffmann$^{1}$, P. N. Best$^{3}$ and T. M. Heckman$^{4}$\\
$^{1}$Max-Planck-Institut f\"{u}r Astrophysik, Karl-Schwarzschild-Str. 1,
Postfach 1317, D-85741 Garching, Germany\\
$^{2}$MPA/SHAO Joint Center for Astrophysical Cosmology at Shanghai Astronomical
Observatory, Nandan Road 80, Shanghai 200030, China\\
$^{3}$SUPA, Institute for Astronomy, Royal Observatory Edinburgh, Blackford
Hill, Edinburgh EH9 3HJ, UK\\
$^{4}$Center for Astrophysical Sciences, Department of Physics and Astronomy,
Johns Hopkins University, Baltimore, MD 21218, USA}

\begin{document}
\pagerange{\pageref{firstpage}--\pageref{lastpage}} \pubyear{2009}
\maketitle
\label{firstpage}

\begin{abstract}
We compute the cross-correlation between a sample of 14,000 radio-loud AGN
(RLAGN) with redshifts between 0.4 and 0.8 selected from the Sloan Digital Sky
Survey and a reference  sample  of 1.2 million luminous red galaxies in the same
redshift range. We quantify how the clustering of radio-loud AGN depends on host
galaxy mass and on radio luminosity. Radio-loud AGN are clustered more strongly
on all scales than control samples of radio-quiet galaxies with the same stellar
masses and redshifts, but the  differences are  largest on scales less than
$\sim 1$~Mpc. In addition, the clustering amplitude of the RLAGN varies
significantly with radio luminosity on scales less than $\sim 1$~Mpc. This
proves that the gaseous environment of a galaxy on the scale of its dark matter
halo, plays a key role in determining not only the probability that a galaxy is
radio-loud AGN, but also the total luminosity of the radio jet. Next, we compare
the clustering of radio galaxies with that of radio-loud quasars in the same
redshift range. Unified models predict that both types of active nuclei should
cluster in the same way. Our data show that most RLAGN are clustered more
strongly than radio-loud QSOs, even when the AGN and QSO samples are matched in
both black hole mass and radio luminosity. Only the most extreme RLAGN and
RLQSOs in our sample, with radio luminosities in excess of
$\sim10^{26}$~W~Hz$^{-1}$, have similar clustering properties. The majority of
the strongly evolving RLAGN population at redshifts $\sim 0.5$ are found in
different environments to the quasars, and hence must be triggered by a
different physical mechanism.
\end{abstract}

\begin{keywords}
galaxies: evolution -- galaxies: active -- radio continuum: galaxies --
quasars: general
\end{keywords}

\section{Introduction}
In recent years, galaxy formation models have become increasingly 
interested in the radio AGN phenomenon, because it is hypothesized that these
objects may regulate the star formation history and mass assembly of the most
massive galaxies and black holes in the Universe. Nearby radio galaxies in
clusters are observed to inject a significant amount of energy into the
surrounding gas. As the radio jets expand and interact with the surrounding
medium, they are believed to heat the gas and prevent further accretion onto
the central galaxy.

The precise conditions that determine whether an AGN develops radio jets/lobes
are still a matter of debate. Several studies have shown that the probability
for a galaxy to become radio-loud is a strong function of stellar mass and
redshift (e.g. \citealt{best05}; \citealt{donoso}). The role that the
environment plays in triggering or regulating the RLAGN phenomenon is not as
well established.

\citet{ledlow} found that the fraction of radio sources and the shape of the
bivariate radio-optical was the same for objects in cluster and field
environments. \citet{best07} found that group and cluster galaxies had similar
radio properties to field galaxies, but the brightest galaxies at the centers of
the groups where more likely to host radio-loud AGN than other galaxies of the
same stellar mass. In the local universe, \citet{mandelbaum} analyzed a large
sample of RLAGN at z$\sim$0.1. They showed that RLAGN inhabit massive dark
matter halos ($>10^{12.5}$~M$_{\odot}$) and that, at fixed stellar mass,
radio-loud AGN are found in more massive dark matter halos than control galaxies
of the same mass selected without regard to AGN properties. This result implies
that RLAGN follow a different halo mass - stellar mass relation than normal
galaxies. \citet{mandelbaum} also found that the boost towards larger halo
masses did not depend on radio luminosity. \citet{hickox} investigated the
clustering in a small sample of higher redshift radio-loud AGN selected from the
AGN and Galaxy Evolution Survey (AGES). They found no difference in the
clustering amplitude of radio galaxies when compared to normal galaxies matched
in redshift, luminosity and color.

Most nearby RLAGN lack any of the standard accretion-related signatures that
would indicate that their black holes are growing significantly at the present
day (\citealt{hardcastle06}). In contrast, quasars are thought to be powered by
supermassive black holes accreting at close to the Eddington rate. Large
redshift surveys like the Two Degree Field Galaxy Redshift Survey (2dFGRS) and
the Sloan Digital Sky Survey (SDSS) now provide angular positions, accurate
photometry and spectra for tens of thousands of QSOs. Recent determinations of
the quasar two-point correlation function have demonstrated that at $z<2.5$
quasars cluster like normal $L_*$ galaxies (\citealt{croom}; \citealt{coil})
and populate dark matter halos of $\sim 10^{12}$~M$_{\odot}$, with the
clustering only weakly dependent on luminosity, color and virial black hole
mass (\citealt{shen1}).

As one moves out in redshift, the radio-loud AGN population evolves very rapidly
in radio luminosity. Whether the RLAGN population also evolves strongly in black
hole accretion rate, is considerably less clear. In particular, our
understanding of whether there is a relationship between powerful, high redshift
radio-loud AGN and quasars is quite sketchy. Around 10\% of the quasar
population is radio-loud. Numerous investigations have found that radio-loud
quasars and at least {\em some} powerful radio galaxies share a number of common
characteristics, such as excess infrared emission, comparable radio morphologies
and luminosities, optical emission lines, large evolutionary rates, and host
galaxies with similar properties. It has thus been tempting to link both
phenomena under the hypothesis that they are the same active nuclei viewed at
different orientations (e.g. \citealt{barthel89}; \citealt{urry}).

A few facts are believed to be key in any attempt to understand the transition
from the population of low-luminosity radio AGN produced by weakly accreting
black holes at low redshifts, to a population of high-luminosity radio AGN that
may be produced by strongly accreting black holes at high redshifts.
\citet{fanaroff} found an important correlation between radio morphology and
radio power: low luminosity sources (Fanaroff-Riley Class I, FRI) show emission
peaking close to the nuclei that fades toward the edges, whereas more luminous
sources (Fanaroff-Riley Class II, FRII) are brightest toward the edges.
\citet{hine} discovered that radio galaxies could also be classified according
to the strength of their optical emission lines: low excitation (weak-lined)
radio galaxies or LERGs, and high excitation (strong-lined) objects or HERGs.
Modern unification models usually associate quasars with the most powerful
HERGs, and low luminosity LERGs with BL Lac objects. Although there is a notable
correspondence between RLAGN luminosity, morphology and spectral type, i.e.
lower luminosity FRIs with LERGs, and higher luminosity  FRIIs with HERGs, the
correlations between these properties are not straightforward. There are
populations of FRI sources with high excitation nuclear lines, and conversely,
FRII galaxies with low excitation spectra are also common.

It has been known for years that very high redshift ($z>2$), powerful radio
galaxies are often surrounded by galaxy overdensities with sizes of a few Mpc
(e.g. \citealt{pentericci}; \citealt{miley}). Since we know that quasars at the
same redshift are clustered like normal $L_*$ galaxies, this would seem to throw
some doubt on a simple unified scheme for explaining both phenomena.

In view of this highly complex situation, a more statistical approach to
comparing the properties of quasars and radio galaxies may yield further
insight. In this paper we present measurements of the projected
cross-correlation between a  sample of 14,000 radio-loud AGN with a median
redshift of $z=0.55$ with the surrounding population of massive galaxies
($M_*>10^{11}$~M$_{\odot}$). The large size of our  samples allows us to
investigate in detail how clustering depends on stellar mass and on radio
luminosity. By comparing the RLAGN clustering with results from control samples
matched in redshift, luminosity and mass, we isolate the effect that the radio
AGN phenomenon has on the clustering signal. We cross-correlate radio quasars
drawn from the SDSS with the same reference sample of massive galaxies. Again,
by using control samples matched in black hole mass and radio luminosity, we
ensure that we compare RLAGN and RLQSOs in as uniform a way as possible.

This paper is organized as follows. In Section 2 we describe the surveys and
samples used in this work. In Section 3 we explain the methodology adopted to
calculate the two-point correlation function. Section 4 presents the results on
radio-loud AGN and quasar clustering. Finally, in Section 5 we summarize our
results and discuss the implications of this work.

Throughout the paper we assume a flat $\Lambda$CDM cosmology, with
$\Omega_{m}=0.3$ and $\Omega_{\Lambda}=0.7$. Unless otherwise stated, we adopt 
$h=H_{0}$/(100 km s$^{-1}$) and present the results in units of Mpc~$h^{-1}$
with $h=1$.

\section{Data}
\subsection{The MegaZ-LRG Galaxy Catalogue}
The Sloan Digital Sky Survey (\citealt{york}; \citealt{stoughton}) is a
five-band photometric and spectroscopic survey that has mapped almost a quarter
of the whole sky, providing precise photometry for more than 200 million objects
and accurate redshifts for about a million galaxies and quasars. The MegaZ-LRG
(\citealt{collist}) is a photometric redshift catalogue based on imaging data
from the fourth Data Release (DR4) of the SDSS. It consists of $\sim$1.2 million
Luminous Red Galaxies (LRG) with limiting magnitude  ${\it i}<20$ over the
redshift range $0.4<z<0.8$. MegaZ adopts various color and magnitude cuts
to isolate red galaxies at $0.4<z<0.8$. The cuts are very similar to those
adopted by the `2dF-SDSS LRG and Quasar' project (2SLAQ, \citealt{cannon}).
Accurate photometric redshifts are available for the entire LRG sample. These
are derived using a neural network photometric redshift estimator (ANNz,
\citealt{collistlah}) that was trained using a sample of $\sim13000$ LRG with
spectroscopic redshifts selected from 2SLAQ. The r.m.s. average photometric
redshift error for all the galaxies in the sample is $\sigma_{rms}=0.049$.

\subsection{The Radio-loud Galaxy Sample}
By combining the optical MegaZ-LRG catalogue with data from the NRAO VLA Sky
Survey (NVSS; \citealt{condon}) and the VLA Faint Images of the Radio Sky at
Twenty Centimeters (FIRST; \citealt{becker}), \cite{donoso} constructed a
catalogue of 14453 radio-loud AGN with 1.4 GHz fluxes above 3.5 mJy. The
cross-matching method utilized a collapsing algorithm to identify
multiple-component FIRST and NVSS sources and the method  was optimized  to take
advantage of both surveys. NVSS has sufficient surface brightness sensitivity to
provide accurate flux measurements of extended radio sources with lobes and
jets. On the other hand, the superior angular resolution of FIRST is crucial to
identify the central core component of each radio source and to provide a
robust association between the radio source and the optically-identified host
galaxy.

Monte-Carlo simulations were used to estimate the reliability ($\sim$98.3\%)
and completeness level (95\%) of the catalogue. The vast majority of the
detected radio AGN (78.6\%) are single component sources in both NVSS and FIRST.
There is, however, a significant fraction of objects without (catalogued) high
S/N FIRST detections ($\sim$8\%), so the authors introduced a method for
analyzing radio maps that allowed them to dig deeper into the FIRST survey and
to use lower S/N detections to pinpoint the location of the host galaxy. We
refer the reader to the original paper by \cite{donoso} for a detailed
description of these procedures and the matching algorithm.

\subsection{The Radio-loud and Radio-quiet Quasar Samples}
In this work we use quasars selected from the fourth edition of the
spectroscopic quasar catalogue (\citealt{schneider}). This contains 77,429
quasars drawn from SDSS DR5, with luminosities larger than $M_i=-22$, that have
at least one broad emission line with FWHM$>$1000~km~s$^{-1}$ in their spectra.
The catalogue also identified radio-loud quasars with FIRST components within a
2~arcsec radius.

Most of the objects targeted as quasars were initially selected using
the algorithm of \citet{richards}, which pick candidates using $ugriz$ broadband
photometry and by matching with unresolved FIRST sources. As the survey
progressed, the quasar selection software was modified to improve its efficiency
at high redshift. This is reflected in two spectroscopic target selection flags
listed as TARGET and BEST (for the final algorithm). Photometry of quasars is
also available in two versions, TARGET measurements (values used at the time of
targeting), and BEST measurements (values derived with the latest pipeline). We
note that the selection of UV-excess quasars at low redshifts ($z<3$) has
remained essentially unchanged, so that only small differences arise from using
TARGET or BEST versions. In addition, the bias introduced by  selection of
targets via FIRST radio detections is significant only at high redshift.

In this work we are interested in cross-correlating the quasars with the LRGs 
described above. We therefore selected an homogeneous quasar sample consisting
of all quasars with $0.35<z<0.78$ and psf magnitudes in the range $15<i<19.1$.
We only consider primary objects ($primary=1$) with point source morphology
($morphology=0$), that were also targeted as primary science objects
($scienceprimary=1$). This yields a sample of 7128 quasars.

Of these 7128 quasars, 684 (9.6\%) have radio identifications in the FIRST
survey down to the 1 mJy flux density limit. One issue that could affect the
derived radio luminosities of the QSOs in our sample is that a fraction of them
present a truly extended FRII-like morphology, and the total radio flux is
distributed over many components. The exclusion of such structures might lead to
an underestimation of the total radio luminosity. We visually examined
NVSS/FIRST radio maps of the 678 QSO with FIRST detections and added the NVSS
fluxes of the associated component(s), if present, or of the FIRST component(s)
when no NVSS source was found in the nearby. For some radio QSO the derived
radio luminosities increase by a factor of $\sim$2-3. Nevertheless, we note
that we repeated the clustering analysis described in Section
\ref{sect:clustering_qso} using only the central (core) component flux, and we
verified that this has no significant influence on any of our results.

According to convention, we define radio-loud quasars as those with total
integrated 1.4~GHz radio power (after adding all associated components) above
$10^{25}$~W~Hz$^{-1}$. With this definition, there are 307 radio-loud quasars in
our sample. We consider objects below this luminosity (or non-detections) as
radio-quiet quasars.

\begin{figure}
\includegraphics[width=84mm]{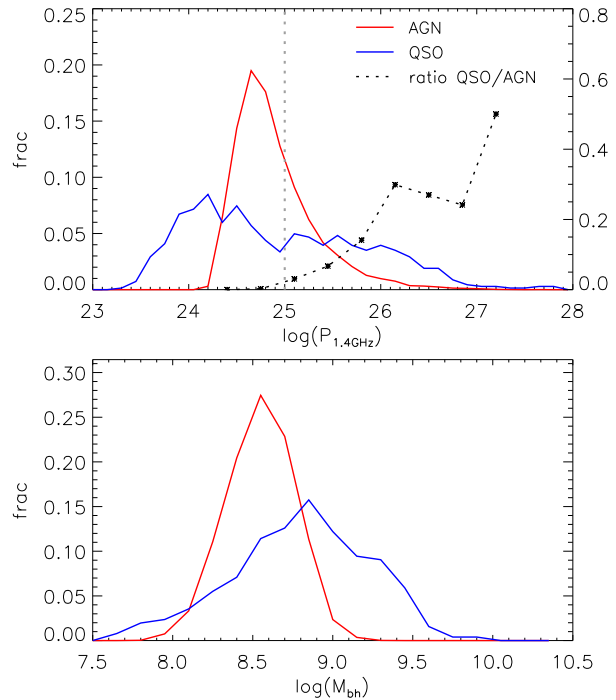}
\caption{Top: normalized distribution of radio luminosity (P$_{\rm1.4 GHz}$)
corresponding to radio-loud AGNs (red), and to QSOs detected down to 1~mJy in
the FIRST survey (blue). The vertical line at 10$^{25}$~W~Hz$^{-1}$ marks the
adopted threshold between radio-quiet and radio-loud QSOs (dotted). Also shown
is the ratio of the number of radio-loud quasars relative to radio AGN (scale on
the right axis). Bottom: distribution of black hole mass (M$_{bh}$) for
radio-loud AGNs and radio-loud QSOs.}
\label{fig:dist_all}
\end{figure}

\subsection{Sample Properties}
\citet{shen2} have derived virial black hole mass estimates for SDSS~DR5
quasars. These are based on H$\beta$, MgII, and CIV emission lines, and the
continuum luminosities around these lines. We adopt these estimates for our
quasar sample (at $z<0.7$, these are mostly derived from H$\beta$). For RLAGN
we adopted the relation between black hole mass-bulge mass derived by
\citet{haring}, M$_{bh}=0.0014$M$_{bulge}$, where we replace M$_{bulge}$ by the
stellar mass of the galaxy. At the lower end of our galaxy mass distribution
($\sim 10^{11} M_{\odot}$), use of the stellar mass instead of the bulge mass
may cause the black hole mass to be overestimated by a factor of $\sim 1.2-1.4$.
We note that the majority of RLAGN in our sample are more massive than this.

For reference, Figure \ref{fig:dist_all} shows the radio luminosity and black
hole mass distributions derived for all the radio-loud AGN and radio-loud QSO in
our samples. In the upper panel, we also plot the ratio of the number of RLQSOs
to RLAGN and show that this increases from $\sim 1\%$ at
$10^{25}$~W~Hz$^{-1}$ up to $\sim50\%$ at $10^{27}$~W~Hz$^{-1}$. This is
broadly consistent with the results of \citet{lawrence} who found similar
relative proportions between broad-lined and narrow-lined 3CR sources.
One scenario that has been introduced to explain this varying fraction (at least
at luminosities where the relative numbers of RLQSO and RLAGN are similar) is
the receeding torus model (e.g. \citealt{simpson}), in which the inner radius of
the obscuring torus (which is identified with the dust sublimation radius)
scales with luminosity as $L^{0.5}$. This model predicts that a larger fraction
of more luminous objects are classified as quasars.

\section{Clustering Analysis}
\subsection{The Cross-correlation Function}
A standard way to characterize the clustering of galaxies is with the two-point
correlation function $\xi(r)$, which measures the excess in the numbers of pairs
with separation $r$ in a volume $dV$, respect to a random distribution with the
same mean number density of objects $n$ (\citealt{peebles}). This can be
expressed as
\begin{equation}
dP = n^{2}[1+\xi(r)]dV^{2}
\end{equation}
Objects are said to be clustered if $\xi>0$. The amplitude and shape of the
correlation function yield a variety of different information. On scales
larger than a few Mpc, the amplitude is a measure of the mass of dark matter
halos that the galaxies inhabit (e.g. \citealt{sheth}). On intermediate scales,
the shape of the correlation function is sensitive to how galaxies are
distributed within their halos (\citealt{li_narrow}), while at scales smaller
than a few hundred kpc it probes processes such as mergers or interactions
(\citealt{li2008}).

Several estimators for the (auto)correlation function have been proposed in the
literature. In this work we calculate the auto-correlation function of the LRGs
using the estimator of \citet{hamilton},
\begin{equation}
\xi(r)=\frac{DD(r)RR(r)}{[DR(r)]^{2}} - 1
\end{equation}
where $DD(r)$, $RR(r)$, $DR(r)$ refers to the normalized number of (LRG-LRG),
(random-random), and (LRG-random) pairs as function of the spatial separation
$r$ (see the next section for details about the construction of the random
sample).

To estimate the cross-correlation function of radio-loud AGN or quasars with the
MegaZ-LRG galaxy sample, we count the number of LRGs around each AGN or quasar
as a function of distance, and divide by the expected number of pairs for a
random distribution,
\begin{equation}
\xi(r)=\frac{CD(r)}{CR(r)} - 1
\end{equation}
where $CD(r)$ stands for the number of (RLAGN/QSO-LRG) pairs, $CR(r)$ is the
number of (RLAGN/QSO-random) pairs, and the quantities have been normalized by
the number of objects in the LRG and random catalogues. The advantage of our
procedure is that it does not require full knowledge of the QSO or RLAGN
selection function. Only the LRG selection function is needed for the
construction of the random sample, and this is well quantified. Another reason
for calculating cross-correlations rather than auto-correlations, is that it
allows us to overcome shot noise when the sample size is small. We note that the
LRG sample ($D$ in the notation above) remains fixed throughout this work. The
error bars of the auto and cross-correlation functions are calculated via
statistical bootstrapping by drawing $n=100$ random samples with replacement.

In practice, photometric redshift errors as well as distortions due to peculiar
velocities along the line of sight will introduce systematic effects in our
estimate of $\xi(r)$. Therefore, to recover real-space clustering properties we
decompose $\xi$ in two directions, along the line of sight ($\pi$) and
perpendicular to it ($r_{p}$). Integrating over the $\pi$-direction allows to
define the projected two-point cross-correlation function $w_{p}(r_{p})$, a
quantity that is independent of such distortions (\citealt{davispeebles}). A
detailed description of the method can be found in \citet{li_narrow}.

\subsection{Construction of the Random Sample}
The random sample used in estimating the cross-correlation function should have
the same selection effects as the observed galaxies. We follow the method by
\citet{li}: we take observed LRG sample inside the coverage mask of SDSS DR4 and
randomly re-assign the sky coordinates of each galaxy. All other quantities such
as like redshift, stellar mass and luminosity are kept fixed. Because the survey
covers a very wide area ($>$6000~deg$^2$ for SDSS DR4), this procedure is
sufficient to remove any coherence in the radial direction and it ensures that
the geometry of the random catalogues are exactly the same as the real one, and
that all redshift-dependent selection effects are accounted for. We generate
N=10 random samples in this way.

\section{Results}
\subsection{Radio-loud AGN clustering}
\begin{figure}
\includegraphics[width=82mm]{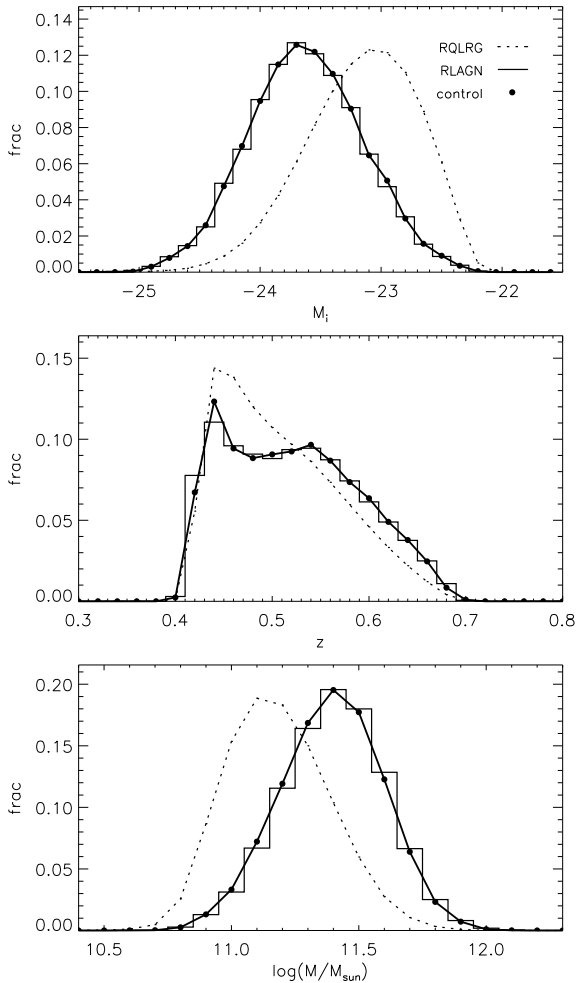}
\caption{Normalized distributions of $i$-band absolute magnitude, redshift and
stellar mass for  radio-quiet LRGs (dotted), radio-loud AGN (histogram), and
control radio-quiet LRGs (large dots).}
\label{fig:contdist_mzrad}
\end{figure}

It is well known that the clustering amplitude of galaxies varies as a function
of mass, luminosity and redshift. Radio AGN are usually hosted by very massive,
$>3L_{*}$ galaxies (\citealt{best05}; \citealt{donoso}). To take this into
account, we select control samples of radio-quiet MegaZ-LRG galaxies with
redshifts, stellar masses and absolute magnitudes that closely match the radio
AGN sample. For each RLAGN we randomly select 10 radio-quiet LRG (or 5,
depending on the number of available candidates) within a tolerance of $\Delta
z=0.02$ in redshift, $\Delta M=0.1$ in log stellar mass, and $\Delta M_{i}=0.05$
in absolute magnitude, where $M_{i}$ is the extinction and k-corrected $i$-band
absolute magnitude. Figure~\ref{fig:contdist_mzrad} shows the distributions of
these parameters for radio-loud, radio-quiet and control objects.

\begin{figure}
\includegraphics[width=82mm]{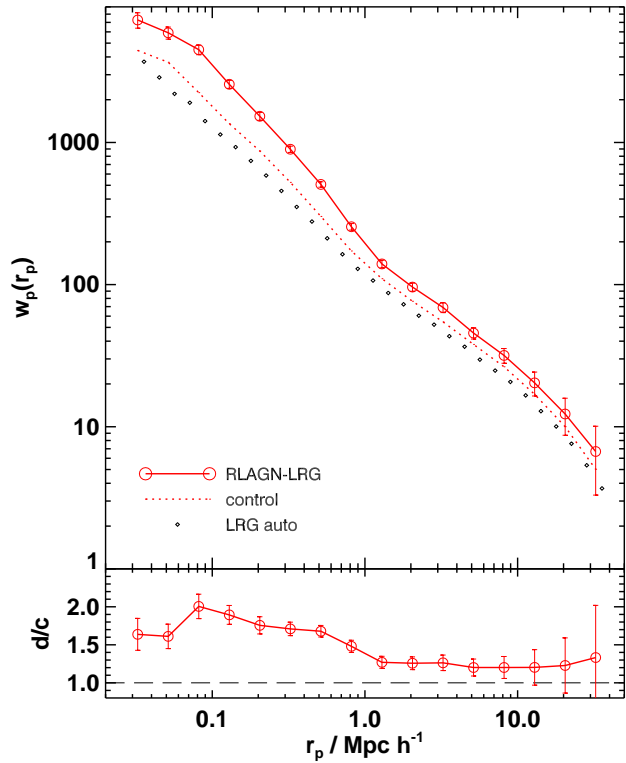}
\caption{Projected cross-correlation function $w_{p}(r_{p})$ between radio-loud
AGN and MegaZ luminous red galaxies (red, solid) in the range 0.3 to 30
Mpc~$h^{-1}$. The LRG-LRG auto-correlation is indicated by small diamond
symbols. Also shown is the cross-correlation of a control sample of radio-quiet
LRG (red, dashed) with the same distribution of redshifts, luminosities and
stellar masses as the radio-loud population. The bottom panel shows the ratio of
$w_{p}(r_{p})$ for the RLAGN to that for the control sample.}
\label{fig:cc_mzrad}
\end{figure}

Using the methods described in the previous section, we first calculate the
auto-correlation function for our reference sample of luminous red galaxies.
We then cross-correlate the radio-loud AGN with the LRG parent sample. This
is shown in Figure~\ref{fig:cc_mzrad}, where it can easily be appreciated that
RLAGN are significantly more clustered than the LRG population on all spatial
scales. The two terms of the clustering signal, corresponding to galaxies within
the same halo (1-halo) and in different halos (2-halo), are clearly visible 
with the transition occurring around 1~Mpc~$h^{-1}$. The boost in clustering
signal exhibited by the RLAGN is significantly stronger on scales less than
1~Mpc~$h^{-1}$, which tells us that RLAGN must occupy special positions within
their dark matter halos. We intend to model this in more detail in upcoming
work.

If we compare the clustering of RLAGN with that of control galaxies with the same
redshifts, luminosities and stellar masses, we see that RLAGN are still
significantly more clustered. The ratio between the cross-correlations
$w_{p}(r_{p})$ corresponding to RLAGN and its corresponding control radio-quiet
sample, is plotted in the bottom panel. This proves that the probability of a
galaxy to become radio-loud depends on environment as well as on black hole or
galaxy mass.

\subsubsection{Dependence on Stellar Mass}
\begin{figure}
\includegraphics[width=82mm]{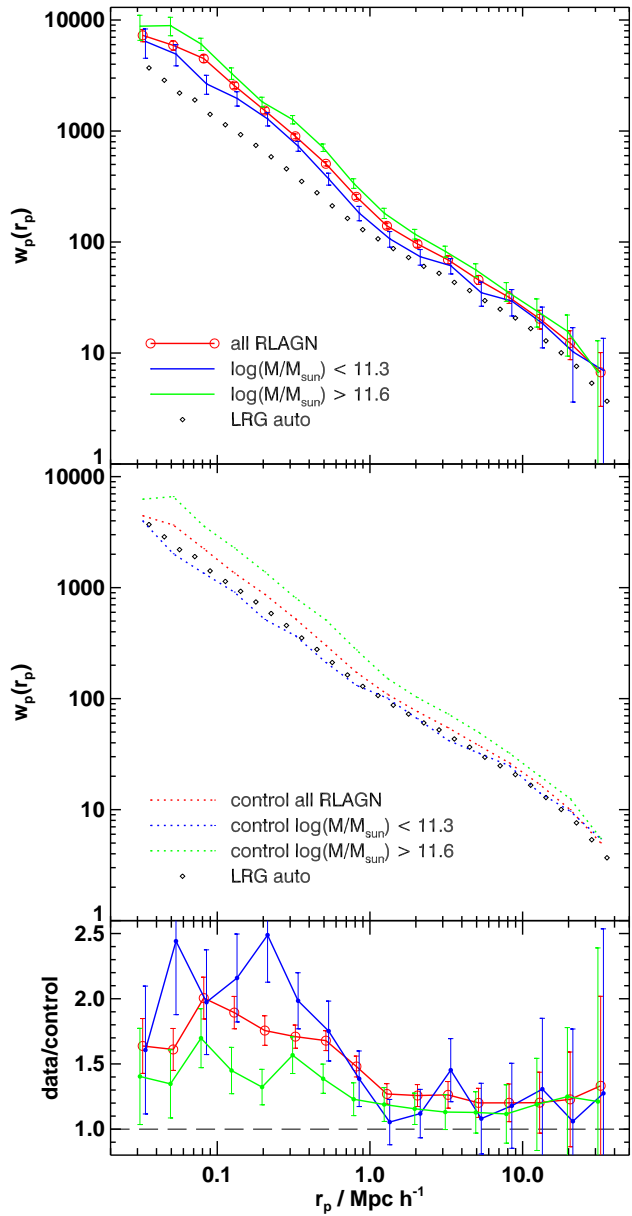}
\caption{Top: projected cross-correlation function $w_{p}(r_{p})$ between
radio-loud AGN and MegaZ luminous red galaxies (red) in the range 0.3 to 30
Mpc~$h^{-1}$. The green and blue lines indicate the cross-correlation of
massive objects with log$(M/M_{\odot})>11.6$, and of less massive systems with
log$(M/M_{\odot})<11.3$. The LRG-LRG auto-correlation function is shown for
reference (diamond symbols). Middle: cross-correlation of control samples of
radio-quiet LRGs that have the same distribution of redshift and stellar mass as
the radio-loud systems. Bottom: ratio of $w_{p}(r_{p})$ between RLAGN and their
corresponding control samples. Note the curves are slightly shifted along the
x-axis to improve the visibility.}
\label{fig:rad_msplit}
\end{figure}

We split the RLAGN sample into two subsamples with log$(M/M_{\odot})<11.3$ and 
with log$(M/M_{\odot})>11.6$. We also applied the same split to the
corresponding control samples. The resulting cross-correlations are plotted in
Figure~\ref{fig:rad_msplit}. As expected, more massive radio galaxies  are more
strongly clustered on all scales. When compared to control galaxies, both
subsamples show roughly the same relative clustering strength on scales larger
than 1-2 Mpc~$h^{-1}$. On small scales the difference between control and data
samples is more significantly boosted for RLAGN in less massive galaxies.
These results are in good agreement with those of \citet{mandelbaum} for RLAGN
at lower redshifts.

\begin{figure}
\includegraphics[width=82mm]{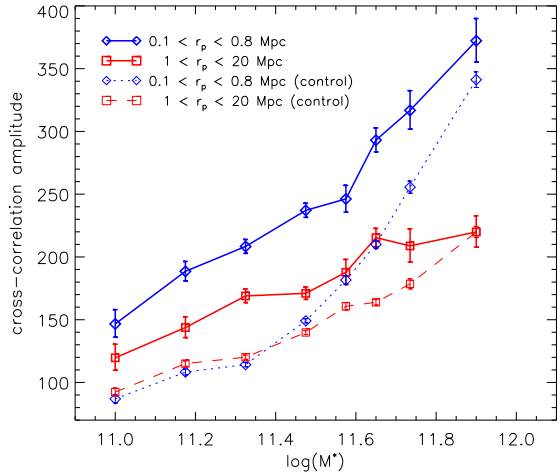}
\caption{The dependence of the cross-correlation amplitude of RLAGN and control
galaxies on stellar mass. The amplitude is computed by fitting a power law with
fixed exponent (see text for details). Results are shown for RLAGN (solid lines)
and their corresponding control radio-quiet LRG (dotted, dashed lines). Fits are
calculated at two different spatial scales, $0.1<r_p<0.8$~Mpc~$h^{-1}$ (blue)
and $1<r_p<20$~Mpc~$h^{-1}$ (red).}
\label{fig:fitmass_ragn}
\end{figure}

We now investigate how the clustering of RLAGN and their control galaxies
varies {\em as a function} of stellar mass. We fit two power laws of the form
$w(r_p)=A~r_{p}^{(1-\gamma)}$ to the cross-correlation function, one over the
range $0.1<r_p<0.8$~Mpc~$h^{-1}$ and the other over the range
$1<r_p<20$~Mpc~$h^{-1}$. This division allows us to quantify separately the
clustering signal contributed by LRGs within the same halo as the RLAGN and by
LRGs residing in different halos. For the complete RLAGN sample, the best
fitting parameters are $A=233.9\pm15$ and $\gamma=2.18\pm0.05$ on scales less
than 1~Mpc~$h^{-1}$ and $A=173.2\pm10$ and $\gamma=1.81\pm0.05$ on larger
scales. We then divide the sample into 8 mass bins and perform new fits, keeping
the slope of the power law fixed and allowing the normalization to vary.
Figure~\ref{fig:fitmass_ragn} shows the  cross-correlation amplitudes 
as a function of stellar mass for  RLAGN and the  radio-quiet control sample. As
can be seen, the ratio between the clustering amplitude of the RLAGN and the
control galaxies depends both on stellar mass and on the scale at which the
clustering is evaluated. On scales less than 1~Mpc~$h^{-1}$, there is a
relatively strong dependence of the ratio on stellar mass, with RLAGN in low
mass galaxies clustered much more strongly than the controls, but RLAGN in
high mass galaxies clustered similarly to the controls. On larger scales, there
is a much weaker trend with mass.

\subsubsection{Dependence on Radio Luminosity}
We now investigate if there is any dependence of RLAGN clustering on the
luminosity of the radio source. \citet{prestage} studied the local galaxy
density around radio galaxies at $z<0.25$, finding that weak FRI sources are
typically found in denser regions compared to the more luminous FRII sources.
\citet{yates} (and later \citealt{hill}) extended such studies to higher
redshifts, concluding that powerful radio galaxies at $z\sim0.5$ are typically
found in environments three times richer than their counterparts at $z\sim0.2$,
but also that the most luminous objects (P$_{\rm178 MHz}\sim
10^{27.1}$~W~Hz$^{-1}$~sr$^{-1}$) occupy richer environments than the weaker
objects (P$_{\rm178 MHz}\sim 10^{26.1}$~W~Hz$^{-1}$~sr$^{-1}$). However, given
the limitations of the samples available, they were unable to determine if such
clustering trends were primarily dependent on redshift or on radio luminosity,
or on a combination of both. \citet{best04} studied the density of galaxies
around nearby radio-loud AGN. He found a positive correlation between local
density and radio luminosity for RLAGN without emission lines, but found that
RLAGN with emission-lines tended to avoid regions of high density.

\begin{figure}
\includegraphics[width=82mm]{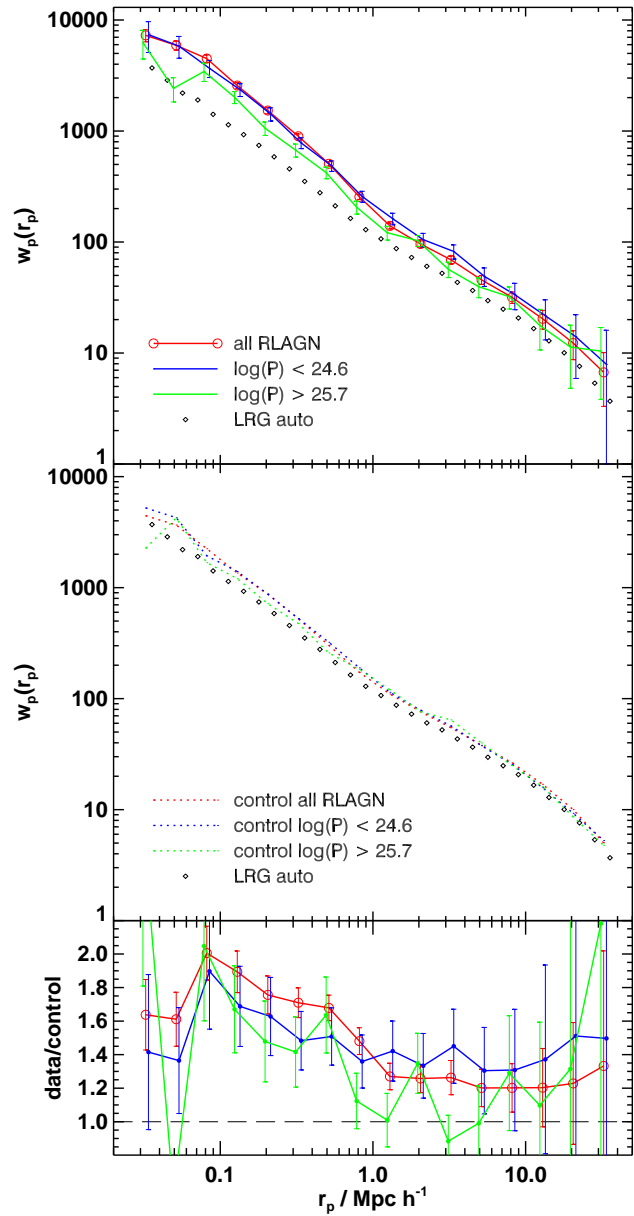}
\caption{Top: projected cross-correlation function $w_{p}(r_{p})$ between
radio-loud AGN and MegaZ luminous red galaxies (red) in the range 0.3 to
30~Mpc~$h^{-1}$. Green and blue lines indicate the cross-correlation of
luminous objects with log(P$_{\rm1.4 GHz}$[W~Hz$^{-1}$])$>25.7$, and of less
powerful AGN with log(P$_{\rm1.4 GHz}$[W~Hz$^{-1}$])$<24.6$. The LRG-LRG
auto-correlation is shown for reference (diamond symbols). Middle:
cross-correlation of control samples of radio-quiet LRGs that have the same
distribution of redshift and stellar mass as the radio-loud systems. Bottom:
ratio of $w_{p}(r_{p})$ between RLAGN and their corresponding control samples.
Note the curves are slightly shifted along the x-axis to improve the
visibility.}
\label{fig:rad_psplit}
\end{figure}

In this study, we split our RLAGN sample into a low-luminosity subsample with 
log(P$_{\rm1.4 GHz}$[W Hz$^{-1}$])$<24.6$, and a high-luminosity subsample
log(P$_{\rm1.4 GHz}$[W Hz$^{-1}$])$>25.7$. These cuts allow us to sample the
faint and bright end  of the radio luminosity distribution. We again build
control samples in the same way as before and we present the cross-correlation
results in Figure~\ref{fig:rad_psplit}. The top panel of
Figure~\ref{fig:rad_psplit} shows that low luminosity RLAGN are more clustered
than high luminosity systems at all scales. When compared to control samples,
this ``boost" in clustering is only visible at scales larger than
$\sim$1~Mpc~$h^{-1}$.

\begin{figure}
\includegraphics[width=82mm]{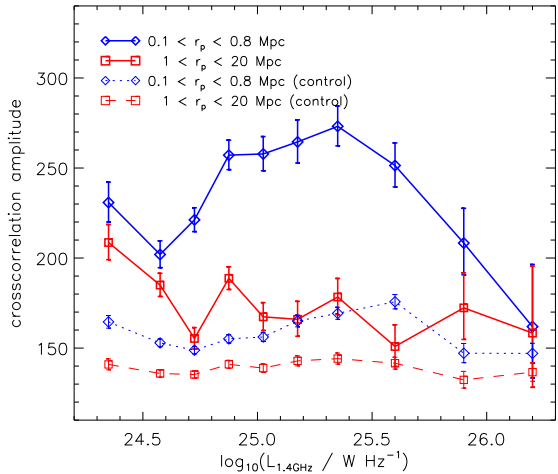}
\caption{The dependence of the cross-correlation amplitude on radio luminosity. 
Results are shown for both, RLAGN (solid lines) and their corresponding control
radio-quiet LRGs (dotted, dashed lines). Fits are calculated for two different
ranges in scale:  $0.1<r_p<0.8$~Mpc~$h^{-1}$ (blue) and
$1<r_p<20$~Mpc~$h^{-1}$ (red).}
\label{fig:fitlogp_ragn}
\end{figure}

To quantify the variation of clustering with radio luminosity in more detail, we
once again proceed by fitting a power law to the cross-correlation functions
for RLAGN subsamples split by radio luminosity. We fit separate power laws
on scales below and above 1~Mpc~$h^{-1}$. The variation in the clustering
amplitude with luminosity is plotted Figure~\ref{fig:fitlogp_ragn}. Two
interesting features can be observed. First, the clustering amplitude of radio
galaxies on large scales is only very weakly  anti-correlated with radio power.
On small scales, the clustering {\em increases} with radio luminosity, peaks at
log(P$_{\rm1.4 GHz}$[W~Hz$^{-1}$])$\sim25.3$, and then decreases for most
luminous radio sources.

\citet{barthel96} argue that the confining effect of a dense intracluster medium
reduces the adiabatic loses of radio lobes, leading to higher levels of
synchrotron emission. Thus, a dense environment may provide a more effective
`working surface' for the lobes, giving rise to the positive correlation between
small-scale clustering amplitude and radio luminosity observed in
Figure~\ref{fig:fitlogp_ragn} for sources with log(P$_{\rm1.4 GHz}$[W
Hz$^{-1}$])$<25.3$. Alternatively, higher radio luminosities in denser
environments may be a result of increased jet powers resulting from the higher
cooling rates in these denser regions. Why does the clustering amplitude drop
for radio sourceswith luminosities higher than this value? As we will argue in
the next section, a radio luminosity of
log(P$_{\rm1.4 GHz}$[W~Hz$^{-1}$])$\sim25.3$ may mark the beginning of a
transition to a population of AGN that are more similar to the quasars, which as
we will show, are significantly less clustered that the RLAGN.

\subsection{Quasar clustering and AGN Unification}\label{sect:clustering_qso}
In this section, we compare the clustering of radio galaxies and quasars at
z$\sim$0.5. Our goal is to develop a better understanding of the relationship
between these two types of active galaxy.

AGN unification models provide an appealing way to account for the diversity of
the observed AGN population. The basic hypothesis is that the observed
characteristics of AGN depend mainly on their orientation relative to the
line-of-sight. Comprehensive reviews of unification models can be found in
\citealt{barthel89}, \citealt{antonucci} or \citealt{urry}.

\begin{figure}
\includegraphics[width=84mm]{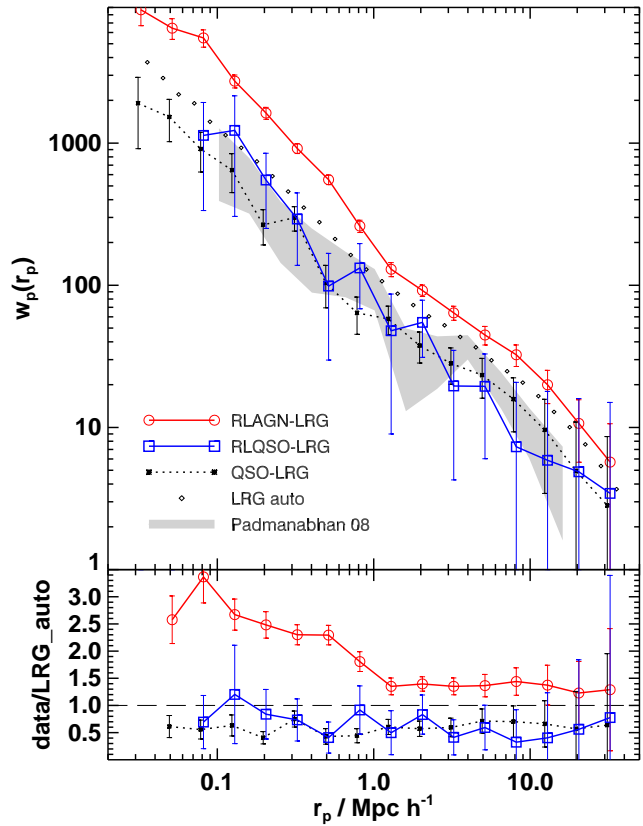}
\caption{Projected cross-correlation function $w_{p}(r_{p})$ between quasars
and LRGs (dotted, black), and between radio-loud quasars and LRGs (blue). For
comparison, we plot the cross-correlation of radio-loud AGN and LRGs (red), as
well as the LRG-LRG auto-correlation (small diamonds). The grey shaded area
indicates the QSO-LRG cross-correlation derived by \citet{padma}. The bottom
panel shows the ratio of $w_{p}(r_{p})$ respect to the LRG auto-correlation. The
analysis is restricted to sources with integrated luminosities above
$10^{25}$~W~Hz$^{-1}$.}
\label{fig:cc_qso}
\end{figure}

In this paper, we attempt to test one fundamental requirement of the unification
scheme of radio-loud objects, namely that the environment of radio
galaxies and radio quasars should be statistically identical. We note that
previous work has already suggested that low excitation radio galaxies (which
include most FRI sources, but also a significant fraction low luminosity FRII
radio galaxies) do not participate in the same unification framework as quasars
or broad line radio galaxies (e.g. \citealt{hardcastle04};
\citealt{hardcastle07}). We will therefore confine our attention to the most
luminous radio-loud galaxies and radio-loud quasars in our sample, i.e. both
with luminosities in excess of $10^{25}$~W~Hz$^{-1}$.

Up to now, observational evidence has not yielded conclusive evidence as to 
whether powerful RLAGN and RLQSOs cluster in the same way. The first problem is
that the available  samples have been small. In the local universe, powerful
radio galaxies with log(P$_{\rm1.4 GHz}$[W Hz$^{-1}$])$\sim26$ have typical
comoving densities of 10$^{-8}$~Mpc$^{-3}$~dex$^{-1}$ at z$\sim$0.1, so large
volumes are required  to detect a significant number of sources. \citet{smith}
studied the environments $\sim$30 low redshift radio quasars and powerful radio
galaxies, concluding that both populations were clustered in much the same way
as radio-quiet QSOs. At higher redshifts ($0.3<z<0.5$), \citet{yates} also found
that the environments of radio galaxies and radio-loud quasars were similar,
with higher luminosity  systems slightly more clustered. \citet{barr} found that
luminous radio-loud quasars exist in a variety of environments including  rich
clusters, compact groups and in low-density environments.

In this work we calculate the cross-correlation function between radio-quiet
and radio-loud quasars, and the same reference sample of LRGs used in Section
4.1. The resulting $w_{p}(r_{p})$ are plotted in Figure \ref{fig:cc_qso}. As can
be seen, {\em there is no significant difference in clustering strength between
radio-loud and radio-quiet quasars}. It is interesting that the clustering
strength of RLAGN seems to be larger than that of radio-loud quasars on  all
scales, and particularly at $r_{p}<$1~Mpc~$h^{-1}$. The mean relative bias of
radio-loud quasars respect to the LRG population remains roughly constant at
$\sim$0.7, while the bias of RLAGN varies strongly from $\sim$2.5 to $\sim$1.5
below 1~Mpc~$h^{-1}$, and then stays relatively constant at larger scales.

We note that our quasar/LRG cross-correlation function agrees extremely well
with that derived by \citet{padma}. \citet{shen1} analyzed the clustering of
radio-loud and radio-quiet quasars in SDSS DR5 at $0.4<z<2.5$ and found that
radio quasars cluster more strongly than radio-quiet quasars with the same black
hole masses. As we will show in section 4.2.1., matching the radio-quiet and
radio-loud quasar sample in black hole mass does not alter our conclusion. We
speculate that disagreement with \citet{shen1} may arise because we consider
a much narrower range in redshift. We note that \citet{wold} also found little
difference between the environments of radio-loud and radio-quiet quasars over
roughly the same redshift range as that probed in this study.

We conclude, therefore, that powerful radio galaxies appear to be hosted by
very massive halos, more massive than their quasar counterparts. In principle,
this suggests that the unification scheme for the two classes of AGN is not as
straightforward as first thought, and additional parameters other than
orientation are required to explain the difference between RLAGN and RLQSOs.

\subsubsection{Black Hole Mass}
One such parameter could  be the mass of the black hole. Some observational
evidence supports the idea that radio jet power might be closely related to the
mass of the black hole and its accretion rate. Links between radio luminosity
and black hole mass have been found in radio galaxies (\citealt{franceschini}),
and in quasars (\citealt{lacy}; \citealt{boroson}). However, other authors have
argued against such strong correlations (\citealt{ho}; \citealt{snellen};
\citealt{metcalf}).

\begin{figure}
\includegraphics[width=82mm]{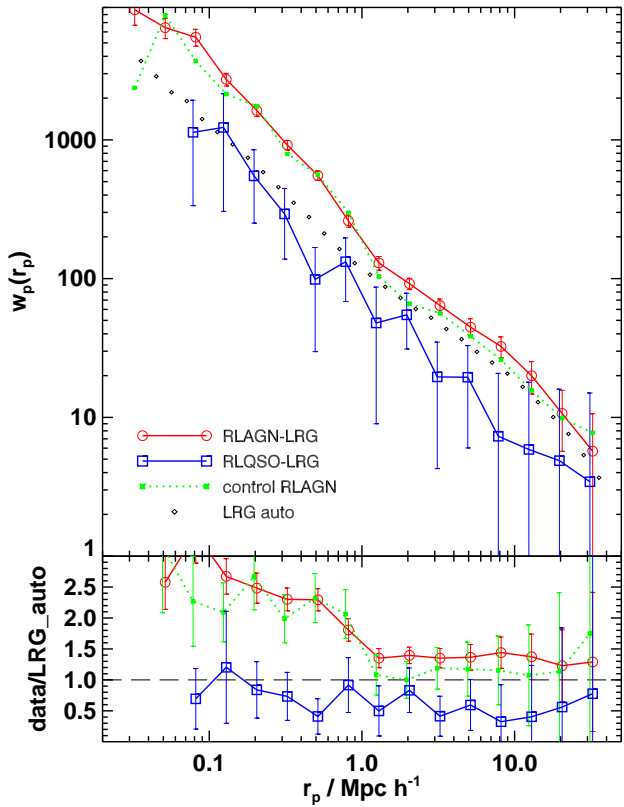}
\caption{Projected cross-correlation function $w_{p}(r_{p})$ between radio-loud
quasars and LRGs (blue). Also shown is the cross-correlation of a control sample
of radio-loud AGN (green) selected to have a similar distribution of
M$_{bh}/$P$_{\rm1.4 GHz}$ as the radio-loud quasars. For comparison we plot the
cross-correlation of RLAGN and LRG (red), and the LRG-LRG auto-correlation
(small diamonds). The bottom panel shows the ratio of $w_{p}(r_{p})$  to the LRG
auto-correlation. The analysis is restricted to sources with integrated
luminosities above $10^{25}$~W~Hz$^{-1}$.}
\label{fig:qso_lrg_MS2}
\end{figure}

\begin{figure}
\includegraphics[width=84mm]{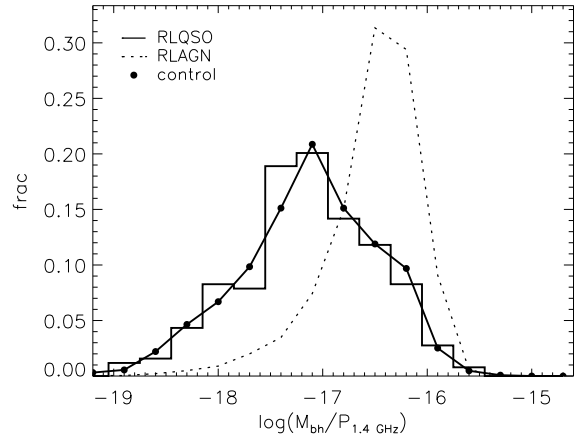}
\caption{Normalized distributions of M$_{bh}/$P$_{\rm1.4 GHz}$ for radio-loud
AGN (dotted), radio-loud QSOs (histogram), and control radio AGN (large dots)
selected to have a similar distribution in M$_{bh}/$P$_{\rm1.4 GHz}$ as the 
radio-loud quasars.}
\label{fig:contdist_qsoms2}
\end{figure}

To control for the effect of black hole mass, we constructed a sample of RLAGN
with a similar distribution in M$_{bh}/$P$_{\rm1.4 GHz}$ as that of radio-loud
quasars. The parameter M$_{bh}/$P$_{\rm1.4 GHz}$ can be considered as a kind of
inverse Eddington ratio that measures how much radio emission per unit black
hole mass is produced by the jet. Figure \ref{fig:qso_lrg_MS2}
shows the resulting cross-correlations. A slight decrease in $w_{p}(r_{p})$
is observable at scales above 1~Mpc~$h^{-1}$ for the RLQSO sample, but the
effect is of low significance. Figure~\ref{fig:contdist_qsoms2} shows the
distribution in M$_{bh}/$P$_{\rm1.4 GHz}$ of the RLAGN and RLQSO samples before
and after the matching procedure.

We find that the difference in clustering  does not change for the samples that
are matched in M$_{bh}/$P$_{\rm1.4 GHz}$, meaning that black hole mass has a
negligible influence in driving the observed differences between the clustering
of radio quasars and radio galaxies. From Figure \ref{fig:dist_all} it can be
seen that some radio-loud quasars are hosted by  black holes more massive than
$10^{9.3}$~M$_{\odot}$, which are not present in the RLAGN population (we
suspect that errors in the virial black hole mass estimates are  to blame). We
have repeated the cross-correlation analysis of radio quasars with black hole
masses in the range $10^{8}$~M$_{\odot}<M_{bh}<10^{9}$~M$_{\odot}$ and find that
this makes no difference to our results.

\subsubsection{Radio Luminosity}

\begin{figure}
\includegraphics[width=82mm]{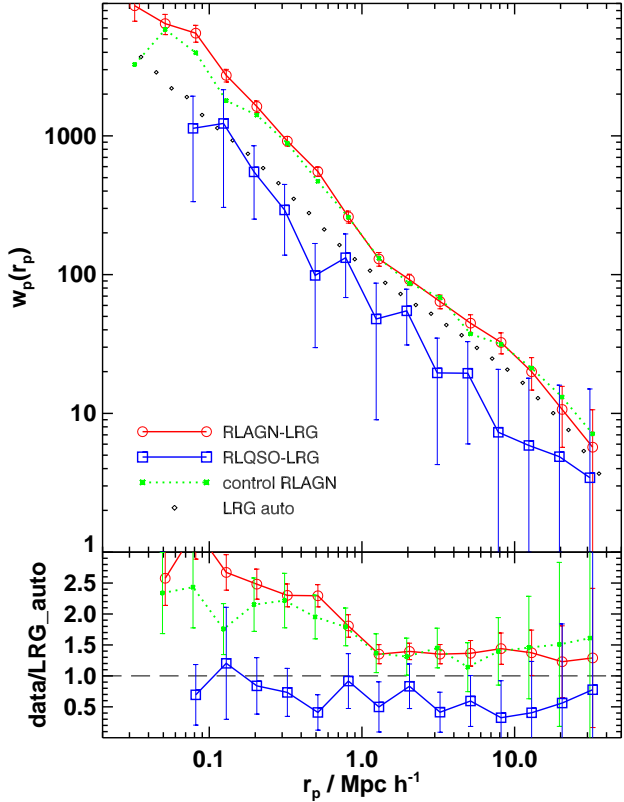}
\caption{Projected cross-correlation function $w_{p}(r_{p})$ between radio
quasars and LRG (blue). Also shown is the cross-correlation of a control sample
of radio-loud LRG (green) selected to have a similar distribution of
log(P$_{\rm1.4 GHz}$) as in radio quasars. For comparison we plot again
cross-correlation of RLAGN and LRG (red), and the LRG-LRG auto-correlation
(small diamonds). The bottom panel shows the ratio of $w_{p}(r_{p})$ respect to
the LRG auto-correlation. The analysis is restricted to sources with integrated
luminosities (after adding all associated components) above
$10^{25}$~W~Hz$^{-1}$.}
\label{fig:qso_lrg_MS1}
\end{figure}

\begin{figure}
\includegraphics[width=84mm]{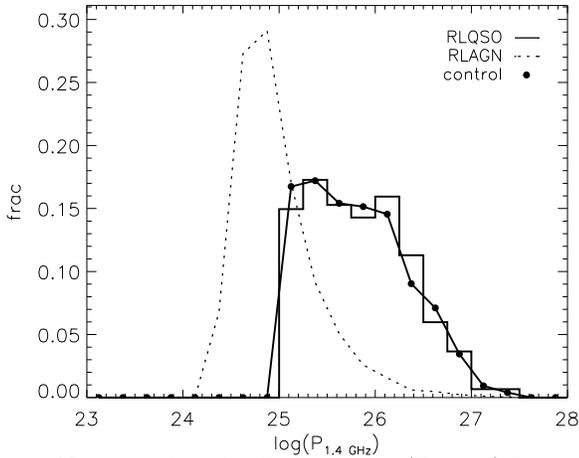}
\caption{Normalized distributions of log(P$_{\rm1.4 GHz})$ for radio-loud AGN
(dotted), radio-loud QSO (histogram), and control radio AGN (large dots)
selected to have a similar distribution as of radio-loud quasars.}
\label{fig:contdist_qsoms1}
\end{figure}

It is also interesting to investigate whether clustering differences
between radio-loud AGN and quasars depend on radio luminosity. To test this, we
build  control samples using the same methodology as before, but this time
matching in log(P$_{\rm1.4 GHz}$)\footnote{We note that a fraction of the
RLQSOs will be core-dominated, so that a fraction of the luminosity of some
sources will be due to beaming. This would affect the matching in radio
luminosity between beamed and non-beamed objects. However, because of the weak
dependence of clustering amplitude on radio luminosity for the quasars, this
effect does not influence our conclusions}. Figures~\ref{fig:qso_lrg_MS1} and
\ref{fig:contdist_qsoms1} show the corresponding cross-correlation functions and
radio luminosity distributions of the matched samples. The clustering of RLAGN
remains essentially unchanged.

\begin{figure}
\includegraphics[width=84mm]{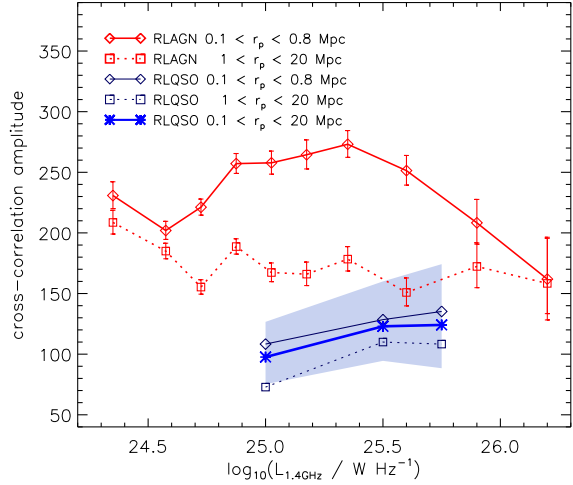}
\caption{Change of the cross-correlation amplitude for RLAGN (red) and RLQSO
(blue), obtained by fitting a power law with varying amplitude and fixed
exponent. Radio quasars are splitted in bins of increasing radio luminosity
[log(P$_{\rm1.4 GHz}$)$>$25.0,$>$25.5,$>$25.75]. Fits are calculated at two
different spatial scales, $0.1<r_p<0.8$~Mpc~$h^{-1}$ (solid) and
$1<r_p<20$~Mpc~$h^{-1}$ (dotted). A single fit over the range
$0.1-20$~Mpc~$h^{-1}$ is indicated by the thick blue line, enclosed by the
shaded error region.}
\label{fig:lrg_qso_comp}
\end{figure}

We now calculate cross-correlation functions for radio-loud quasars of
increasing radio luminosities [log(P$_{\rm1.4 GHz}$)$>$25.0,$>$25.5,$>$25.75].
We do find an increase in clustering strength as function of radio power on all
scales in the range $0.1<r_p<20$~Mpc~$h^{-1}$. This is plotted in
Figure~\ref{fig:lrg_qso_comp}, where we compare the cross-correlation amplitude
of radio-loud AGN and QSO. The amplitude is calculated using a single power-law
fit over the entire range, since the correlation function of quasars does not
exhibit a clear break at a scale of $\sim$1~Mpc~$h^{-1}$, as is the case for
radio galaxies. We find that RLAGN are  more strongly clustered than RLQSO at
all radio luminosities that we are able to probe. However, the clustering
amplitude of RLQSOs increases as a function of radio luminosity, while that of
RLAGNs decreases at the very highest radio luminosities. Extrapolation of our
results suggests that both kinds of AGN might have similar clustering at radio
luminosities in excess of $10^{26}$~W~Hz$^{-1}$. This would imply that the
unified model for radio-loud quasars and radio galaxies can only be valid at
these very high radio luminosities. This is consistent with the dependence of
the relative numbers of the two AGN types as a function of radio power (Figure
\ref{fig:dist_all}).

\section{Summary}
In this work, we have successfully applied cross-correlation techniques to
characterize the  environments of $\sim 14,000$ radio-loud AGN with
P$_{\rm1.4 GHz}>10^{24}$~W~Hz$^{-1}$, selected from $\sim 1.2$ million LRG at
$0.4<z<0.8$. We have also compared the clustering of RLAGN with that of
radio-loud quasars over the same redshift interval. By using control samples of
radio-quiet objects matched in redshift, stellar mass and optical luminosity
(or radio luminosity, when appropriate) we have isolated the effect such
parameters have in influencing the clustering signal. The main results of this
paper can be summarized as follows:

\begin{itemize}
\item
Radio AGN at $0.4<z<0.8$ are substantially more clustered than their parent
luminous red galaxy population. Radio-loud AGN are also more strongly clustered
than radio-quiet galaxies of the same stellar mass and redshift. The clustering
differences are largest on scales less than 1~Mpc~$h^{-1}$.
\item
Radio-loud AGN hosted by more massive galaxies are more strongly clustered than
those hosted by less massive galaxies. However, the clustering {\em difference}
between RLAGN and control samples of radio-quiet galaxies is most pronounced
for RLAGN in low mass hosts.
\item
We study the dependence of the clustering amplitude on the luminosity 
of the radio source. For $r_p>1$~Mpc~$h^{-1}$ there is a weak, but significant
anti-correlation with radio power. For $r_p<1$~Mpc~$h^{-1}$ the dependence of
clustering amplitude on luminosity is more complex: the cross-correlation
amplitude increases with luminosity up to $\sim10^{25.3}$~W~Hz$^{-1}$, and then
decreases for the most luminous radio sources in our sample.
\item
We have compared the environments of radio-loud AGN and radio-loud QSOs.
RLAGN are clustered more strongly than RLQSOs on all scales, indicate that they
populate dark matter halos of different mass. These results hold even when the
RLAGN and RLQSO samples are matched in radio luminosity and black hole mass.
\item
There are indications that the very most luminous RLAGN and RLQSOs in our
sample (P$>10^{26}$~W~Hz$^{-1}$) do have similar clustering amplitudes. Only at
these very high radio powers are the space-densities of radio-loud quasars
and radio galaxies similar. This implies that unification of the two AGN
populations can only be valid above P$\sim 10^{26}$~W~Hz$^{-1}$.
\end{itemize}

One major limitation of this study with regard to constraining AGN unification
scenarios, is that it is based purely on photometric data from the SDSS, so we
are unable to split our RLAGN sample into high-excitation and low-excitation
sources. It is quite possible that the presence or absence of emission lines
will provide the best way to define a population of radio galaxies that are
clearly unified with the quasars. In this case, we would expect to find that the
high-excitation radio galaxy population would cluster in a similar way to the
quasars.

In addition, we note that because the parent sample of our RLAGN catalogue
consists of luminous {\em red} galaxies, it is also likely that we completely 
miss some number of RLAGN with bluer colors and stronger emission lines. The
analysis of the RLAGN luminosity function presented in \citet{donoso}
indicates that the missing sources cannot constitute more than $\sim$20\% of
the total RLAGN population, so will not dominate the clustering signal of the
radio AGN  population as a whole. Nevertheless the quasar analogues among the
radio galaxy population may still be under-represented in our analysis.

Fortunately, upcoming large spectroscopic surveys such as BOSS will target
nearly complete samples of more than a million massive galaxies at $0.4<z<0.8$
and will provide optical spectra for tens of thousands of  radio galaxies. We
will then be able to quantify the fraction of RLAGN of given radio luminosity
that have emission lines and how the clustering depends on emission line
strength.

The most definitive result to emerge from our analysis is clear proof that
the environment of a galaxy on the scale of the dark matter halo in which
it resides (i.e. on scales of $\sim$1~Mpc~$h^{-1}$ and below), does play a key
role in determining not only the probability that a galaxy a is radio-loud AGN,
but also  the total luminosity of the radio jet. Combining our results with
those of \citet{best05}, we conclude that both black hole mass and environment
must determine the radio-loud character of an active galaxy.

Our previous work also demonstrated that strong evolution of the radio AGN
population only occurs above a characteristic radio luminosity of
$\sim10^{25}$~W~Hz$^{-1}$ (\citealt{donoso}). It is very intriguing that the
results in this paper indicate that this luminosity marks the break point in
clustering trends, and that the radio luminosity where denser environment
ceases to have a boosting influence also is of order $10^{25}$~W~Hz$^{-1}$.

Finally, the strong evolution of the radio source population at  radio
luminosities above $\sim10^{25}$~W~Hz$^{-1}$ combined with the strong
clustering of this population, must imply that the heating rate of the gas in
groups and clusters of galaxies is higher at redshifts $\sim 0.5$ than it is at
the present day. We intend to quantify this in more detail in upcoming work.

\section*{Acknowledgments}
We would like to thank the Max Planck Society for the financial support provided
through its Max Planck Research School on Astrophysics PhD program. PNB is
grateful for support from the Leverhulme Trust.

Funding for the SDSS and SDSS-II has been provided by the Alfred P. Sloan
Foundation, the Participating Institutions, the National Science Foundation, the
US Department of Energy, the National Aeronautics and Space Administration, the
Japanese Monbukagakusho, the Max Planck Society and the Higher Education Funding
Council for England. The SDSS Web Site is http://www.sdss.org/. The SDSS is
managed by the Astrophysical Research Consortium for the Participating
Institutions. The Participating Institutions are the American Museum of Natural
History, Astrophysical Institute Potsdam, University of Basel, Cambridge
University, Case Western Reserve University, University of Chicago, Drexel
University, Fermilab, the Institute for Advanced Study, the Japan Participation
Group, Johns Hopkins University, the Joint Institute for Nuclear Astrophysics,
the Kavli Institute for Particle Astrophysics and Cosmology, the Korean
Scientist Group, the Chinese Academy of Sciences (LAMOST), Los Alamos National
Laboratory, the Max-Planck-Institute for Astronomy (MPA), the
Max-Planck-Institute for Astrophysics (MPIA), New Mexico State University, Ohio
State University, University of Pittsburgh, University of Portsmouth, Princeton
University, the United States Naval Observatory and the University of
Washington.

This research project uses the NVSS and FIRST radio surveys, carried out using
the National Radio Astronomy Observatory Very Large Array. NRAO is operated by
Associated Universities Inc., under cooperative agreement with the National
Science Foundation.

\label{lastpage}

\begin{thebibliography}{99}
\bibitem[\protect\citeauthoryear{Antonucci}{1993}]{antonucci} Antonucci R.,
1993, ARA\&A, 31, 473
\bibitem[\protect\citeauthoryear{Barr et~al.}{2003}]{barr} Barr J. M., Bremer
M. N., Baker J. C., Lehnert M. D., 2003, MNRAS, 346, 229
\bibitem[\protect\citeauthoryear{Barthel}{1989}]{barthel89} Barthel P. D.,
1989, ApJ, 336, 606
\bibitem[\protect\citeauthoryear{Barthel \& Arnaud}{1996}]{barthel96} Barthel
P. D., Arnaud K. A., 1996, MNRAS, 283, 45
\bibitem[\protect\citeauthoryear{Becker et~al.}{1995}]{becker} Becker R. H.,
White R. L., Helfand D. J., 1995, ApJ, 450, 559
\bibitem[\protect\citeauthoryear{Best}{2004}]{best04} Best, P. N., 2004, MNRAS,
351, 70
\bibitem[\protect\citeauthoryear{Best et~al.}{2005}]{best05} Best, P. N.,
Kauffmann G., Heckman T. M., Brinchmann J., Charlot S., Ivezi\'{c} \v{Z}., White
S. D. M., 2005, MNRAS, 362, 25
\bibitem[\protect\citeauthoryear{Best et~al.}{2007}]{best07} Best P. N., von der
Linden A., Kauffmann G., Heckman T. M., Kaiser C. R., 2007, MNRAS, 379, 894
\bibitem[\protect\citeauthoryear{Boroson}{2002}]{boroson} Boroson T. A., 2002,
ApJ, 565, 78
\bibitem[\protect\citeauthoryear{Cannon et~al.}{2006}]{cannon} Cannon R. D. et
al., 2006, MNRAS, 372, 425
\bibitem[\protect\citeauthoryear{Coil et~al.}{2007}]{coil} Coil A. L., Hennawi
J. F., Newman J. A., Cooper M. C., Davis M., 2007, ApJ, 654, 115
\bibitem[\protect\citeauthoryear{Collister \& Lahav}{2004}]{collistlah}
Collister A. A., Lahav O., 2004, PASP, 116, 345
\bibitem[\protect\citeauthoryear{Collister et~al.}{2007}]{collist} Collister A.
et al., 2007, MNRAS, 365, 68
\bibitem[\protect\citeauthoryear{Condon et~al.}{1998}]{condon} Condon J. J.,
Cotton W. D., Greisen E. W., Yin Q. F., Perley R. A., Taylor G. B., Broderick J.
J., 1998, AJ, 115, 1693
\bibitem[\protect\citeauthoryear{Croom et~al.}{2005}]{croom} Croom S. M. et al.,
2005, MNRAS, 356, 415
\bibitem[\protect\citeauthoryear{Davis \& Peebles}{1983}]{davispeebles}
Davis M., Peebles P. J. E., 1983, ApJ, 267, 465
\bibitem[\protect\citeauthoryear{Donoso et~al.}{2009}]{donoso} Donoso E., Best,
P. N., Kauffmann, G., 2009, MNRAS, 392, 617
\bibitem[\protect\citeauthoryear{Fanaroff \& Riley}{1974}]{fanaroff} Fanaroff B.
L., Riley J. M., 1974, MNRAS, 167, 31
\bibitem[\protect\citeauthoryear{Franceschini et~al.}{1998}]{franceschini}
Franceschini A., Vercellone S., Fabian A. C., 1998, MNRAS, 297, 817
\bibitem[\protect\citeauthoryear{Hamilton}{1993}]{hamilton} Hamilton A. J. S.,
1993, ApJ, 417, 19
\bibitem[\protect\citeauthoryear{Hardcastle}{2004}]{hardcastle04} Hardcastle
M. J., 2004, A\&A, 414, 927
\bibitem[\protect\citeauthoryear{Hardcastle et~al.}{2006}]{hardcastle06}
Hardcastle M. J., Evans D. A., Croston, J. H., 2006, MNRAS, 370, 1893
\bibitem[\protect\citeauthoryear{Hardcastle et~al.}{2007}]{hardcastle07}
Hardcastle M. J., Evans D. A., Croston, J. H., 2007, MNRAS, 376, 1849
\bibitem[\protect\citeauthoryear{H\"{a}ring \& Rix}{2004}]{haring} H\"{a}ring
N., Rix H. W., 2004, ApJ, 604, 89
\bibitem[\protect\citeauthoryear{Hickox et~al.}{2009}]{hickox} Hickox R. C. et
al., 2009, ApJ, 696, 891
\bibitem[\protect\citeauthoryear{Hill \& Lilly}{1991}]{hill} Hill G. J., Lilly
S. J., 1991, AJ, 367, 1
\bibitem[\protect\citeauthoryear{Hine \& Longair}{1979}]{hine} Hine R. G.,
Longair M. S., 1979, MNRAS, 188, 111
\bibitem[\protect\citeauthoryear{Ho}{2002}]{ho} Ho L. C., 2002, ApJ, 564, 120
\bibitem[\protect\citeauthoryear{Lacy et~al.}{2001}]{lacy} Lacy M.,
Laurent-Muehleisen S. A., Ridgway S. E., Becker, R. H., White R. L., 2001, ApJ,
551, 17
\bibitem[\protect\citeauthoryear{Lawrence}{1991}]{lawrence} Lawrence A., 1991,
MNRAS, 252, 586
\bibitem[\protect\citeauthoryear{Ledlow \& Owen}{1996}]{ledlow} Ledlow M.
J., Owen F. N., 1996, AJ, 112, 9
\bibitem[\protect\citeauthoryear{Li et~al.}{2006a}]{li} Li C., Kauffmann G.,
Jing Y. P., White S. D. M., B\"{o}rner G., Cheng F. Z., MNRAS, 2006a, 368, 21
\bibitem[\protect\citeauthoryear{Li et~al.}{2006b}]{li_narrow} Li C.,
Kauffmann G., Wang L., White S. D. M., Heckman T. M., Jing Y. P., 2006b, MNRAS,
373, 457
\bibitem[\protect\citeauthoryear{Li et~al.}{2008}]{li2008} Li C., Kauffmann
G., Heckman T. M., Jing Y. P., White S. D. M., MNRAS, 385, 1093
\bibitem[\protect\citeauthoryear{Mandelbaum et~al.}{2009}]{mandelbaum}
Mandelbaum R., Li C., Kauffmann G., White S. D. M., 2009, MNRAS, 393, 377
\bibitem[\protect\citeauthoryear{Metcalf \& Magliocchetti}{2006}]{metcalf}
Metacalf R. B., Magliocchetti M., 2006, MNRAS, 365, 101
\bibitem[\protect\citeauthoryear{Miley et~al.}{2006}]{miley} Miley G. K. et
al., 2006, ApJ, 650, 29
\bibitem[\protect\citeauthoryear{Padmanabhan et~al.}{2008}]{padma} Padmanabhan
N., White M., Norberg P., Porciani C., 2008, MNRAS, 397, 1862
\bibitem[\protect\citeauthoryear{Peebles}{1980}]{peebles} Peebles P. J. E.,
1980, The large-scale structure of the universe. Research supported by the
National Science Foundation. Princeton, N.J., Princeton Universtity Press.
\bibitem[\protect\citeauthoryear{Pentericci et~al.}{2000}]{pentericci}
Pentericci L., 2000, A\&A, 361, 25
\bibitem[\protect\citeauthoryear{Prestage \& Peacock}{1988}]{prestage}
Prestage R. M., Peacock J. A., 1988, MNRAS, 230, 131
\bibitem[\protect\citeauthoryear{Richards et~al.}{2002}]{richards} Richards G.
T. et al., 2002, AJ, 123, 2945
\bibitem[\protect\citeauthoryear{Schneider et~al.}{2007}]{schneider} Schneider
D. P. et al., 2007, AJ, 134, 102
\bibitem[\protect\citeauthoryear{Shen et~al.}{2008}]{shen2} Shen Y., Greene
J. E., Strauss M. A., Richards G. T., Schneider D.P., 2008, ApJ, 680, 169
\bibitem[\protect\citeauthoryear{Shen et~al.}{2009}]{shen1} Shen Y. et al.,
2009, ApJ, 697, 1656
\bibitem[\protect\citeauthoryear{Sheth \& Tormen}{1999}]{sheth} Sheth R. K.,
Tormen G., 1999, MNRAS, 308, 119
\bibitem[\protect\citeauthoryear{Simpson}{1998}]{simpson} Simpson C., 1998,
MNRAS, 297, 39
\bibitem[\protect\citeauthoryear{Smith \& Heckman}{1990}]{smith} Smith E. P.,
Heckman T.M., 1990, AJ, 348, 38
\bibitem[\protect\citeauthoryear{Snellen et~al.}{2003}]{snellen} Snellen I.
A. G., Lehnert M. D., Bremer M. N., Schilizzi R. T., 2003, MNRAS, 342, 889
\bibitem[\protect\citeauthoryear{Stoughton et~al.}{2002}]{stoughton} Stoughton
C. et al., 2002, AJ, 123, 485
\bibitem[\protect\citeauthoryear{Urry \& Padovani}{1995}]{urry} Urry M. C.,
Padovani P., 1995, PASP, 107, 803
\bibitem[\protect\citeauthoryear{Yates et~al.}{1989}]{yates} Yates M. G., Miller
L., Peacock J. A.,1989, MNRAS, 240, 129
\bibitem[\protect\citeauthoryear{York et~al.}{2000}]{york} York D. G. et al.,
2000, AJ, 120, 1579
\bibitem[\protect\citeauthoryear{Wold et~al.}{2000}]{wold} Wold M., Lacy M.,
Lilje P. B., Serjeant S., 2000, MNRAS, 316, 267
\end{thebibliography}
\end{document}